# Amplified Dispersive Fourier-Transform Imaging for Ultrafast Displacement Sensing and Barcode Reading

Keisuke Goda, Kevin K. Tsia, and Bahram Jalali

*Department of Electrical Engineering, University of California, Los Angeles, CA 90095*

Dispersive Fourier transformation is a powerful technique in which the spectrum of an optical pulse is mapped into a time-domain waveform using chromatic dispersion. It replaces a diffraction grating and detector array with a dispersive fiber and single photodetector. This simplifies the system and, more importantly, enables fast real-time measurements. Here we describe a novel ultrafast barcode reader and displacement sensor that employs internally-amplified dispersive Fourier transformation. This technique amplifies and simultaneously maps the spectrally encoded barcode into a temporal waveform. It achieves a record acquisition speed of 25 MHz -- four orders of magnitude faster than the current state-of-the-art.

A barcode is a machine-readable binary representation of information which appears as a printed pattern of high and low reflectivity (light and dark) bars. Barcodes represent unique identification tags and are read by a laser or camera scanner that illuminates the pattern with laser light (or incoherent light in the case of camera scanners) and measures reflections with a photodiode or CCD array. Because of their simplicity and ease of scanning, barcodes have become an indispensable tool for parcel labeling, sorting, inventory management, supply chain tracking, and mail distribution [1]. Beyond the common product identification in retail stores, they have other important applications. For example, barcode technology is employed in blood banks to trace the collection, processing, storage, and transfusion of donated blood [2]. Networked readers can collect parcel data from multiple sources and compare them against databases that store information related to the basic product data, tracking history, and processing instructions. These networks can also include a control system that makes decisions and initiate action based on product identity and information from databases and other sensors.

As the volume of information encoded into barcodes is increasing and they become integrated into real-time sensor networks, there is a growing demand for high-speed barcode readers. Similarly, there is a need for high speed scanners for non-contact position and displacement sensing, such as those used for real-time inspection and monitoring in industrial applications. The scan rate of conventional barcode readers and position and displacement sensors is limited to rates on the order of 100 – 10,000 scans per second [3] chiefly by the refresh rate of the CCD or the mechanical scanning speed of laser scanners. The technology described in this paper operates based on a radically different approach, and the proof-of-concept demonstration achieves a scan rate of 25 MHz -- more than 10,000 times higher than what is achievable today. The core technology is capable of achieving ever higher speed in the future.

Exploiting the mathematical equivalence between paraxial diffraction and temporal dispersion [4], dispersive Fourier transformation (FT) is a powerful technique in which the spectrum of a pulse is mapped into a time-domain waveform using group-velocity dispersion. It replaces a diffraction grating and detector array with a dispersive fiber and single photodetector, greatly simplifying the system. It has been used for fast real-time absorption and Raman spectroscopy [5–7], optical coherence tomography [8], and ultrafast analog-to-digital conversion via the photonic time-stretch process [9]. Here we propose and demonstrate a novel imaging method using dispersive FT for ultrafast reflectivity measurement. This method, which we refer to as chirped wavelength encoding and electronic time-domain sampling (CWEETS), exposes a sample to a broadband pulsed beam which is spatially dispersed and diffracted by a diffraction grating so that each separated wavelength of the probe beam is incident on a different part of the sample along a transverse line on the sample. In other words, a single line of different wavelengths is projected onto the sample. The reflectivity profile of the sample is encoded into the spectrum of the reflected light from the sample [10]. The key feature of our method is the utilization of dispersive FT using group-velocity dispersion to decode the spectrally encoded reflectivity profile of the sample in the time domain. This enables ultrafast real-time imaging at an image acquisition rate given by the repetition rate of the laser pulse train (typically on the order of 10 MHz). Equally important, we perform distributed Raman post-amplification of a weak signal within the dispersive element, and simultaneously with the dispersive FT process. This internal amplification overcomes the otherwise detrimental optical loss in the dispersive element associated with extrinsic propagation losses as well as the fundamental loss imposed by the Kramers-Kronig relations [4, 5, 11, 12]. It can even produce net gain, resulting in greatly enhanced sensitivity, acquisition speed, and resolution. Losses can also be overcome by using a high power laser source; however, the Raman post-amplification is advantageous because it avoids potential damage to the sample and unwanted nonlinear signal distortion.

In dispersive FT with CWEETS, the desirable features for a dispersive element are high total dispersion, low loss, large optical bandwidth, smooth dispersion over the bandwidth, and commercial availability. Dispersion compensation fiber (DCF) offers an optimum combination of these parameters and is our preferred choice. While the loss in the

DCF can also be compensated by discrete optical amplifiers such as erbium-doped fiber amplifiers or even semiconductor optical amplifiers, distributed Raman amplification within the dispersive DCF is superior because it maintains a relatively constant signal level throughout the dispersive FT process. This important property maximizes the signal-to-noise-and-distortion ratio by keeping the signal power away from low power (noisy) and high power (nonlinear) regimes. Incidentally, this advantage of distributed Raman amplification over discrete amplification is known in long haul fiber optic communication links [13]. In addition to a lower noise figure, the intrinsic properties of Raman amplification are also favorable for the dispersive FT technique, such as its widely tunable gain spectrum and its naturally broadband gain spectrum allowed by the amorphous nature of optical glass fiber. The gain bandwidth can be further broadened by using multi-wavelength pump lasers, and, surprisingly but fortuitously, extremely broadband gain spectra can be realized using incoherent pump sources [5]. This is highly desirable because a large optical bandwidth results in a large number of resolvable points in the dispersive FT imaging. Raman-amplified dispersive elements also eliminate the need for a high power optical source.

To prove the functionality of the dispersive FT imaging with CWEETS, we constructed an apparatus as shown in Figure 1. The optical source is a mode-locked femtosecond fiber laser with a center wavelength of 1560 nm, a bandwidth of 12.3 nm, a repetition rate of 25 MHz (after pulse picking), and an average power of 5 mW. The pulse train from the optical source is collimated by a fiber collimator and expanded by a beam expander to form a probe beam. This beam is spatially dispersed and diffracted by a diffraction grating with a groove density of 1200 lines/mm and is collimated by an objective lens (NA = 0.85) onto a sample. Since the dispersed beam is projected onto the sample, the reflectivity profile of the sample is encoded into the spectrum of the back-reflected beam. As shown in Figure 2, the sample consists of two patterned aluminum thin films straddled upon a silicon rib waveguide with silicon dioxide over-cladding. It is mounted on a piezo-electric transducer (PZT) which is scanned with a ramp function at 5 kHz. The number of resolvable points is found to be 65 based on the groove density of the grating, the center wavelength, bandwidth, and diameter of the probe beam, and Littrow's angle [10]. A DCF module with a dispersion of -1512 ps/nm temporally disperses and maps the spectrally encoded reflectivity profile into a temporal waveform, which is captured by a single photodetector with a response time of 50 ps and digitized by a digital oscilloscope (Tektronix) with a bandwidth of 16 GHz and a sampling rate of 50 GS/s. Based on the repetition rate of the laser (25 MHz), the repetitive scans are performed every 40 ns. To compensate for the inevitable optical loss (9.5 dB) in the DCF, distributed Raman amplification is simultaneously implemented within the dispersive element by pumping the DCF with two continuous-wave diode lasers at 1470 nm and 1480 nm via wavelength division multiplexers (WDMs) such that the gain bandwidth of the Raman amplifier covers the bandwidth of the spectrally encoded back-reflected light. The Raman pumps used to pump the DCF module are diode lasers (Furukawa) designed for distributed Raman amplification in telecommunications systems. The Raman amplification raises the strength of a weak signal, resulting in improved detection sensitivity.

Figure 2 shows the single-shot calibrated reflectivity profile of the sample with the calibrated length axis above the figure at four different times when the PZT was scanned with a ramp function. Only 14 ns was required to capture the reflectivity profile of the sample. The region of the silicon waveguide transmits most of the probe beam whereas the aluminum films reflect it back toward the optical source. Although the image acquisition period was 40 ns, only one out of every 625 scans (one scan every 25 $\mu$s) is plotted for clarity. The fast displacement of the sample was clearly captured.

Figure 3 shows improved sensitivity of the dispersive FT imaging with CWEETS enabled by distributed Raman post-amplification of a weak reflection signal from the sample at various pump powers. The Raman amplification not only compensates for the inevitable optical loss in the DCF, but also gives net gain, resulting in improved detection sensitivity and capturing the otherwise invisible reflectivity profile.

In another experiment in which the objective lens was replaced with a spherical lens ($f$ = 100 mm) and the beam expansion ratio was also increased to increase the number of resolvable points to 145, we measured the single-shot reflectivity profile of a test barcode with a known code pattern as shown in the background of Figure 4. The test barcode was produced by printing black bars on a transparent film. The figure also shows the calibrated reflectivity profile of the measured barcode with the calibrated length axis above the figure. It clearly shows a matched barcode pattern of 1001010100 captured only within 17 ns.

In summary, we have proposed and demonstrated ultrafast one-dimensional imaging using dispersive FT with CWEETS. This method is ideal for ultrafast real-time displacement sensing and barcode reading.  A record image acquisition speed of 25 MHz, four orders of magnitude improvement over conventional displacement sensors and barcode readers, was demonstrated. In this proof-of-principle demonstration, the bandwidth of the optical circulator and the gain bandwidth of the Raman amplifier (~15 nm) limited the number of resolvable points on the sample. However, this is not an inherent limitation of the technique. The use of optics with larger bandwidth and additional Raman pumps at different wavelengths can significantly increase the optical bandwidth, and hence, the number of resolvable points.

This work was partially supported by DARPA. We are grateful to Shalabh Gupta and Daniel R. Solli at UCLA for valuable discussions.

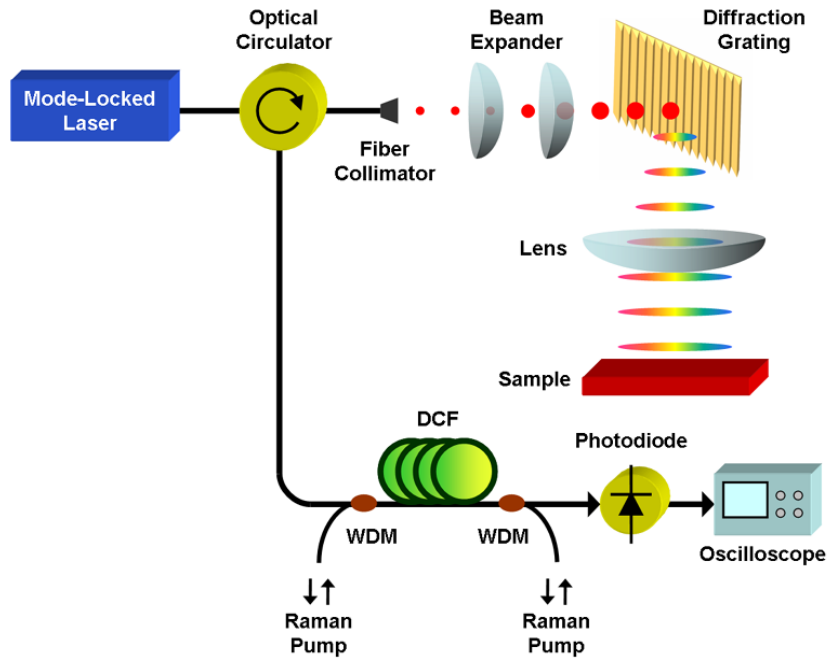

**FIGURE 1**: Schematic of the dispersive Fourier-transform imaging experiment based on amplified dispersive Fourier transformation. DCF: dispersion compensation fiber, and WDM: wavelength division multiplexer. The Raman-amplified DCF maps the spectrally encoded reflectivity profile of the sample into a temporal waveform, which is then captured by the photodetector and digitized by the oscilloscope.

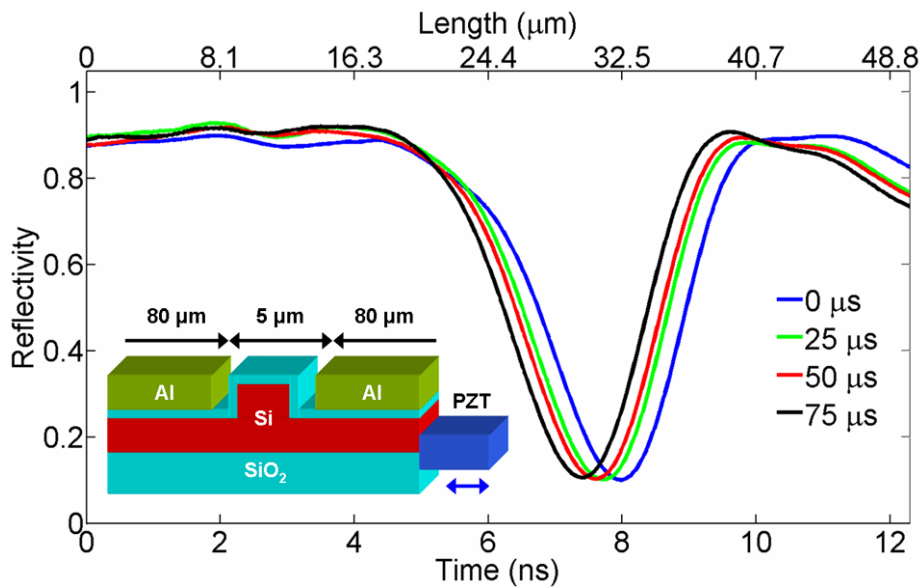

**FIGURE 2**: Calibrated reflectivity profile of the sample (shown in the inset drawing) with the calibrated length axis above the figure. The sample is scanned by the piezoelectric transducer (PZT) with a ramp function. One end of the PZT is attached to the sample while the other end is anchored to a heavy mount. The fast displacement of the sample from the right to the left was clearly captured. Although the image acquisition period is 40 ns, only one out of every 625 scans (one scan every 25 μs) is plotted for clarity.

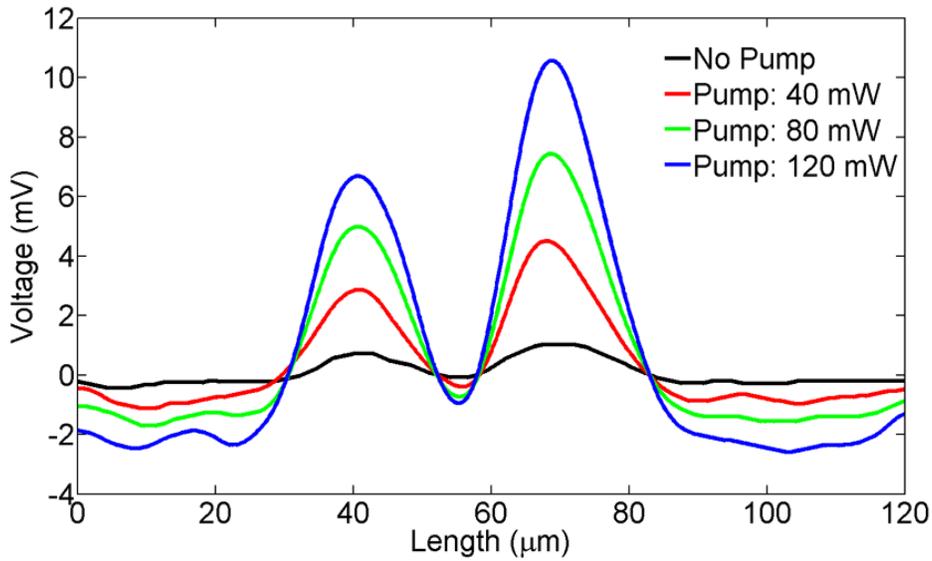

**FIGURE 3**: Improved sensitivity of the dispersive Fourier-transform imaging method enabled by distributed Raman post-amplification of a weak reflection signal from the sample at various pump powers. The otherwise invisible reflectivity profile is improved by the Raman amplification. The x-axis is calibrated to be length whereas the y-axis is the uncalibrated voltage measured by the photodetector.

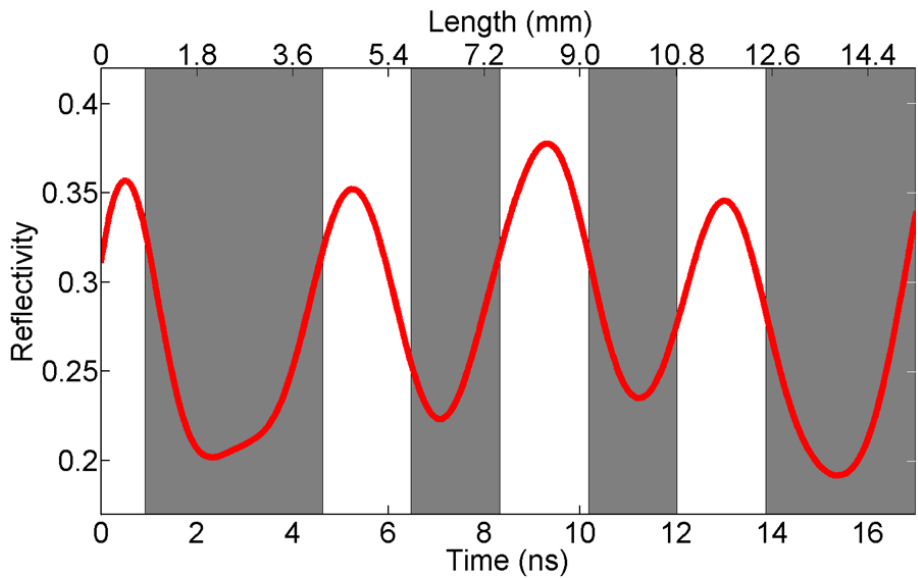

**FIGURE 4**: Calibrated reflectivity profile of a test barcode with its known code pattern in the background and the calibrated length axis above the figure. The figure clearly shows a matched barcode pattern of 1001010100 captured within only 17 ns.